\documentclass[%
 aip,
 amsmath,amssymb,
 reprint,%
]{revtex4-1}

\usepackage{graphicx}
\usepackage{dcolumn}
\usepackage{bm}

\usepackage[utf8]{inputenc}
\usepackage[T1]{fontenc}
\usepackage{mathptmx}
\usepackage{etoolbox}
\usepackage{float} 
\usepackage{titlesec}
\makeatletter
\def\@email#1#2{%
 \endgroup
 \patchcmd{\titleblock@produce}
  {\frontmatter@RRAPformat}
  {\frontmatter@RRAPformat{\produce@RRAP{*#1\href{mailto:#2}{#2}}}\frontmatter@RRAPformat}
  {}{}
}%
\makeatother
\begin{document}

\preprint{AIP/123-QED}

\title[Characteristic of AC and DC driven PHMR sensor]{Dual-mode operation of ring-shaped spin Hall magnetoresistance sensor with biaxial sensing capability\\}
\title[ ]{Dual-mode operation of ring-shaped spin Hall magnetoresistance sensor with biaxial sensing capability\\}
\author{Jiayi Xu}
\affiliation{ 
Department of Electrical and Computer Engineering, National University of Singapore, 4 Engineering Drive 3, Singapore 117583}
\affiliation{
National University of Singapore Guangzhou Research Translation and Innovation Institute, Block B5, No.10, China-Singapore Smart 3rd Street, China-Singapore Guangzhou Knowledge City, Huangpu District, Guangzhou, Guangdong, China}

\author{Yuxin Si}
\author{Jiaqi Wang}
\author{Tingxuan Zhang}
\affiliation{ 
Department of Electrical and Computer Engineering, National University of Singapore, 4 Engineering Drive 3, Singapore 117583}

\author{Zhenfei Hou}
\affiliation{Key Laboratory for Physical Electronics and Devices of the Ministry of Education, School of Electronic Science and Engineering, Faculty of Electronic and Information Engineering, Xi’an Jiaotong University, Xi’an 71004, China }

\author{Yihong Wu$^*$}
\affiliation{ 
Department of Electrical and Computer Engineering, National University of Singapore, 4 Engineering Drive 3, Singapore 117583}%
\email{elewuyh@nus.edu.sg}

\date{\today}

\begin{abstract}
We present a spin Hall magnetoresistance sensor based on a NiFe/Pt multiring bridge structure, which exhibits high sensitivity and good linearity in two perpendicular directions within the sensor plane. Under DC excitation, it responds linearly to magnetic field perpendicular to the current direction, whereas AC excitation enables a linear response with near-zero offset to magnetic field aligned with the current direction, driven by spin-orbit torque effect. Moreover, the AC excitation effectively mitigates low-frequency $1/f$ noise down to sub-$\mu\text{V}/\sqrt{\text{Hz}}$ at 1 Hz. Systematic investigations have been performed to optimize the NiFe thickness while keeping the Pt thickness at 2 nm. The biaxial sensing capability offers a promising approach for multidimensional magnetic field detection in advanced sensing applications.

\end{abstract}

\maketitle

In recent years, magnetoresistive (MR) sensors have attracted widespread attention due to their high sensitivity and ultralow magnetic field detection capability.\cite{ripka2010,lenz2006,diaz2009,wang2008,graham2004,zheng2019,leitao2024,tumanski2013review,heremans1993solid} They have been widely applied in various cutting-edge fields, such as internet of things,\cite{zheng2019} aerospace,\cite{diaz2009} navigation,\cite{tumanski2013review, heremans1993solid} and healthcare. \cite{lim2022advances, wang2008, van1996magnetic} Among the different types of MR sensors, including anisotropic magnetoresistance (AMR),\cite{mcguire1975anisotropic} giant magnetoresistance (GMR),\cite{jeng2014vector, daughton1994gmr} and tunneling magnetoresistance (TMR), the AMR sensor is structurally the simplest as it does not involve a multilayer structure.\cite{chappert2007spin,parkin2003spintronic,wu2004nano,trinh2017miniature} Despite its small output signal, the AMR sensor exhibits the lowest noise among the various types of MR sensors, thus offering the highest detectivity at the single-device level.\cite{stutzke2005,Deak2017} To enhance the output of AMR sensors, planar Hall magnetoresistance (PHMR) sensors have been developed.\cite{schuhl1995low, pham2018highly, roy2020development, ejsing2004planar,tumanski2013review} These sensors utilize a multiring structure that offers a significantly longer current path than conventional rectangular element sensors.\cite{Henriksen2010, oh2011hybrid,lee2021bridge} It is worth noting that the PHMR is fundamentally the same as the AMR, since the output signal still originates from the electrical potential drop along the current path. In parallel, recent advances in spin orbit torque (SOT) technology have further improved the performance of the AMR sensor by integrating SOT and spin Hall magnetoresistance (SMR) into the sensor design.\cite{xu2018ultrathin, Xu2019, xu2017macro, yang2017semitransparent, tsymbal2019spintronics} This has led to the development of SMR sensors that feature near-zero DC offset, negligible hysteresis, and low noise, with their superior performances demonstrated in several proof-of-concept applications.\cite{Xu2020,Lu2021a,Lu2021b,Wu2021}\\
\indent In this work, we incorporate the SOT into a PHMR sensor by replacing the ferromagnetic (FM) layer with an FM/heavy metal (HM) bilayer (hereafter referred to as ring-shaped SMR sesnor as the readout signal is dominantly from the SMR). We systematically examine the sensor’s magnetic response and noise characteristics under both DC and AC excitations. The ring-shaped SMR sensor is shown to exhibit biaxial sensing capability. Under DC excitation, the sensor shows high sensitivity and linearity to magnetic fields perpendicular to the current direction. In contrast, when driven by an AC bias, it provides a robust response with near-zero offset to magnetic field aligned with the current direction, while effectively suppressing low-frequency $1/f$ noise. This orthogonal magnetic response under distinct excitation modes enables dual-axis field sensing within a compact, single-device architecture, offering a promising solution for multidimensional magnetic field detection in advanced sensing applications.
\begin{figure}[htbp]
\centering
\includegraphics[width=\linewidth]{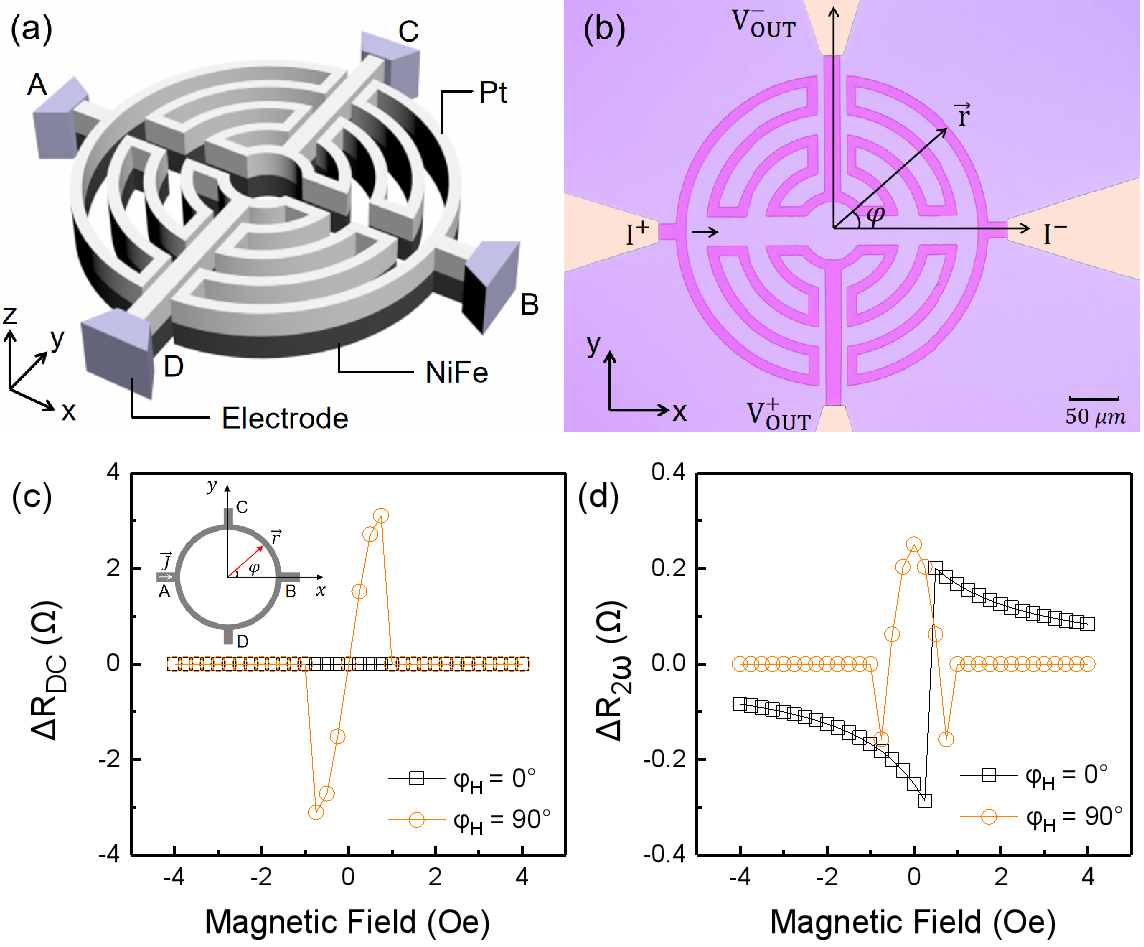}
\caption{\label{fig1} (a) Schematic of the multiring sensor; (b) Optical micrograph of the fabricated sensor. Scale bar is 50 $\mu$m; (c) and (d): simulated response curves of the Pt (2~nm)/NiFe (2.6~nm) sensor under DC (c) and AC (d) excitations, respectively. In both figures, open squares denote the response for \( \varphi_H = 0^\circ \), and open circles correspond to \( \varphi_H = 90^\circ \). The inset of Fig.1(c) shows the schematic of single ring-shaped sensor used for simulation. 
}
\end{figure}

Figure~\ref{fig1}(a) shows the schematic of the sensor which adopts a NiFe($t_{\mathrm{NiFe}}$)/Pt($t_{\mathrm{Pt}}$) bilayer multiring structure comprising five concentric rings, where $t_{\mathrm{NiFe}}$ and $t_{\mathrm{Pt}}$ represent the thicknesses of the NiFe and Pt layers, respectively. The outer radius of the largest ring is 160 $\mu$m, and the ring width is 10 $\mu$m, as shown in Figs.~\ref{fig1}(a) and \ref{fig1}(b). The NiFe thickness has been varied from 2 nm to 2.8 nm. A too small thickness results in weakening and eventual disappearance of the sensor output signal, whereas too large a thickness leads to nonlinearity and large noise. Therefore, we fixed the Pt thickness at 2 nm and varied the NiFe thickness from 2.2 nm to 2.6 nm to examine how the sensor performs within this thickness range. Although increasing the number of rings and reducing the ring width can enhance the output signal by extending the effective current path on the same footprint, it also increases the power consumption. Therefore, these parameters must be carefully optimized in the actual sensor design to balance sensitivity, signal amplitude, and power consumption. The NiFe/Pt bilayers were deposited on SiO\textsubscript{2}/Si substrates, with the NiFe layer deposited first by e-beam evaporation at a rate of 0.2~\AA/s, followed by the deposition of Pt using DC magnetron sputtering at a sputtering power of 50~W (without substrate heating). Both layers were deposited in a multi-chamber system at a base pressure below $3 \times 10^{-8}$~Torr without breaking the vacuum. An in-plane magnetic field of approximately 500~Oe was applied during deposition to induce a weak uniaxial anisotropy in the magnetic film. The bilayers were subsequently patterned into a multiring structure using standard photolithography and lift-off techniques.
Figure~\ref{fig1}(b) shows the optical micrograph of the completed sensor.

To understand how the ring-shaped SMR sensor works under both DC and AC excitation, we first simulate the magnetic response based on a simplified design: a single-ring structure composed of a NiFe ($t_{\mathrm{NiFe}}$) / Pt ($t_{\mathrm{Pt}}$) bilayer, patterned with four electrical contacts labeled A, B, C, and D, equally spaced along the circumference of the ring (see the inset of Fig.\ref{fig1}(c)). Contacts A and B are positioned along the $x$-axis, while contacts C and D are aligned along the $y$-axis. An electrical current is applied between terminals A and B, and the voltage response is measured across terminals C and D to monitor changes induced by the magnetic field. When an external magnetic field is applied to the bilayer, the equilibrium magnetization direction of NiFe, $\varphi_M$, can be determined by minimizing the total magnetic energy density:\cite{coey2010magnetism}
\begin{equation}
\varepsilon = -H M_s t_{\text{NiFe}} \cos (\varphi_m - \varphi_H) + K_u t_{\text{NiFe}} \sin^2 (\varphi_m - \varphi_u)
\end{equation}
where $H$ is the external magnetic field, $M_s$ is the saturation magnetization, $K_u$ is the uniaxial anisotropy constant, $\varphi_H$ is the field angle, and $\varphi_u$ is the anisotropy axis. The first term is the Zeeman energy (aligning $\varphi_m$ with $\varphi_H$), whereas the second term is the uniaxial anisotropy energy (aligning $\varphi_m$ with $\varphi_u$). 

The applied current flows through both the NiFe and Pt layers, with the ratio determined by their respective thicknesses and electrical resistivity. The current in the NiFe layer, $I_{\mathrm{NiFe}}$, contributes to the AMR signal, whereas the current in the Pt layer, $I_{\mathrm{Pt}}$, gives rise to both SOT and SMR effects. For NiFe/Pt bilayers with small thicknesses, we have previously shown that the SMR contribution dominates over AMR;\cite{xu2018ultrathin} therefore, we refer to it as an SMR sensor. As the AMR and SMR exhibit the same in-plane angular dependence on the magnetization direction, the electric field inside the NiFe layer can be expressed as (ignoring anomalous Hall and thermoelectric effects):\cite{harder2016electrical}
\begin{equation}
\vec{E} = \rho_\perp \vec{J} + \Delta \rho_{\text{MR}} (\vec{J} \cdot \vec{m}) \vec{m}
\end{equation}
where $\rho_\perp$ is the resistivity perpendicular to $\vec{m}$, $\vec{J}$ is the effective current density, $\vec{m} = (\cos \varphi_m, \sin \varphi_m)$ is the magnetization unit vector,  and $\Delta \rho_{\text{MR}}$ is the resistivity change due to both AMR and SMR. The effective current density $\vec{J}$ is given by $\vec{J} = \frac{I_0}{2 w t_{\text{eff}}} \hat{\varphi}$ for path AD ($-\hat{\varphi}$ for path CA),  where $\hat{\varphi}=(-\sin \varphi, \cos \varphi)$ and $t_{\text{eff}} = \frac{t_{\text{NiFe}} t_{\text{Pt}}}{t_{\text{NiFe}} + \left( \frac{\rho_{\text{NiFe}}}{\rho_{\text{Pt}}} \right) t_{\text{Pt}}}$.\\

The voltage between terminals C and D can be obtained by integrating the electric field along path CD, i.e.,  $V_{CD}=\int_{C}^{D} \vec{E}\cdot d\vec{r}$. In the ideal case, the drop in electric potential along the CA and AD path cancels out, and therefore the resistance change $\Delta R = V_{CD}/I$ can be expressed as 

\begin{align}
\Delta R &= \frac{r\Delta \rho_{\text{MR}}}{w t_{\text{eff}}} \left( -\int_{\pi/2}^{\pi} \sin^2 (\varphi_m - \varphi) \, d\varphi \right. \notag \\
         &\quad \left. + \int_{\pi}^{3\pi/2} \sin^2 (\varphi_m - \varphi) \, d\varphi \right).
\end{align}

\noindent This corresponds to either a DC resistance change $\Delta R_{DC}$ under DC current excitation or the first harmonic resistance change $\Delta R_{\omega}$ when the sensor is driven by an AC current. In the latter case, the current generates both an Oersted field and a field-like (FL) SOT effective field in the radial direction of the ring (both fields point in the same direction in NiFe/Pt bilayers). Since both fields are proportional to the current, they induce a time-varying modulation of the magnetization, which in turn causes a time-dependent change in the resistance along the current path. This oscillating resistance, when multiplied by the driving AC current, gives rise to a voltage component at twice the driving frequency, resulting in a measurable second harmonic resistance change, given by

\begin{equation}
\begin{split}
\Delta R_{2\omega} = \frac{r \Delta \rho_{\text{MR}}}{w t_{\text{eff}}} \Bigg[ 
    -\int_{\pi/2}^{\pi} \sin 2(\varphi_m - \varphi) \Delta \varphi_{\text{CA}} \, d\varphi \\
    + \int_{\pi}^{3\pi/2} \sin 2(\varphi_m - \varphi) \Delta \varphi_{\text{AD}} \, d\varphi 
\Bigg],
\end{split}
\end{equation}

where:
\begin{subequations}
\begin{equation}
\Delta \varphi_{\text{CA}} = \frac{-H_{\text{FL, Oe}} \cos (\varphi_m + \varphi)}{H_k \cos 2(\varphi_m - \varphi_u) + H \cos (\varphi_m - \varphi_H)},
\end{equation}
\begin{equation}
\Delta \varphi_{\text{AD}} = \frac{H_{\text{FL, Oe}} \cos (\varphi_m + \varphi)}{H_k \cos 2(\varphi_m - \varphi_u) + H \cos (\varphi_m - \varphi_H)}.
\end{equation}
\end{subequations}

\noindent The terms $\Delta \varphi_{\text{CA}}$ and $\Delta \varphi_{\text{AD}}$ represent the perturbation in the magnetization angle due to the combined effects of the SOT and Oersted field along the CA and AD paths, $H_{FL,Oe}=\alpha I/2$, where $\alpha$ is a proportional coefficient, and $H_K = \frac{2K_u}{M_s}$ is the anisotropy field. 


The unique angular dependence of \( \Delta R_{\text{DC}} \) and \( \Delta R_{2\omega} \) enables dual-mode operation of the sensor, as demonstrated in the simulation results shown in Fig.~\ref{fig1}c and Fig.~\ref{fig1}d. The parameters used in the simulation are as follows: \( M_s = 800 \, \text{emu/cm}^3 \), \cite{xu2017macro}\, \( K_u = 4 \times 10^2 \, \text{erg/cm}^3 \), \( t_{\text{NiFe}} = 2.4 \, \text{nm} \), \( t_{\text{Pt}} = 2\, \text{nm} \), \( r = 155 \, \mu\text{m} \) (average radius), \( w = 10 \, \mu\text{m} \) (ring width), \( \rho_{\text{NiFe}} = 20 \times 10^{-6} \, \Omega\cdot\text{cm} \), \( \rho_{\text{Pt}} = 10.6 \times 10^{-6} \, \Omega\cdot\text{cm} \), \( \Delta\rho_{\text{MR}} = 1 \times 10^{-8} \, \Omega\cdot\text{cm} \), $\varphi_u=0^\circ$, \( I = 3 \, \text{mA} \), and \( \alpha = 0.04 \, \text{Oe/mA} \). \cite{xu2017macro}\, The resistivity and magnetoresistance related parameters are within the range of experimental values. The anisotropy is chosen such that, with the assistance of a time-varying SOT effective field, the hysteresis is effectively diminished.\cite{xu2018ultrathin} Figure~\ref{fig1}(c) shows the calculated \( \Delta R_{\text{DC}} \) as a function of the external magnetic field applied along the \( x \)- and \( y \)-directions, respectively. It can be seen that the response is in good linearity when the field is applied along the \( y \)-direction.  In contrast, as shown in Fig.~\ref{fig1}(d), \( \Delta R_{2\omega} \) shows a linear response along the \( x \)-direction and a nonlinear one along the \( y \)-direction. A slight shift of the curve in the positive field direction is due to the numerical accuracy of the simulation, which does not affect the response of the actual sensor. These results clearly demonstrate the dual-mode operation and biaxial sensing capability of the ring-shaped SMR sensor. Although the simulation was performed on single-ring structure, the results shall apply to multi-ring design as well. 

 The performance of the fabricated multiring SMR sensor was characterized in a magnetically shielded environment, including its magnetic response, noise characteristics, and low-frequency field-sensing capability. The magnetic response was evaluated by applying either a DC or AC current between terminals A and B while monitoring the corresponding voltage across terminals C and D. The noise spectral density, along with the minimum detectable magnetic field under both excitation modes, was measured using a Keysight 35670A dynamic signal analyzer without subtracting the electronic noise of the preamplifier.


\begin{figure}[htbp]
\centering
\includegraphics[width=\linewidth]{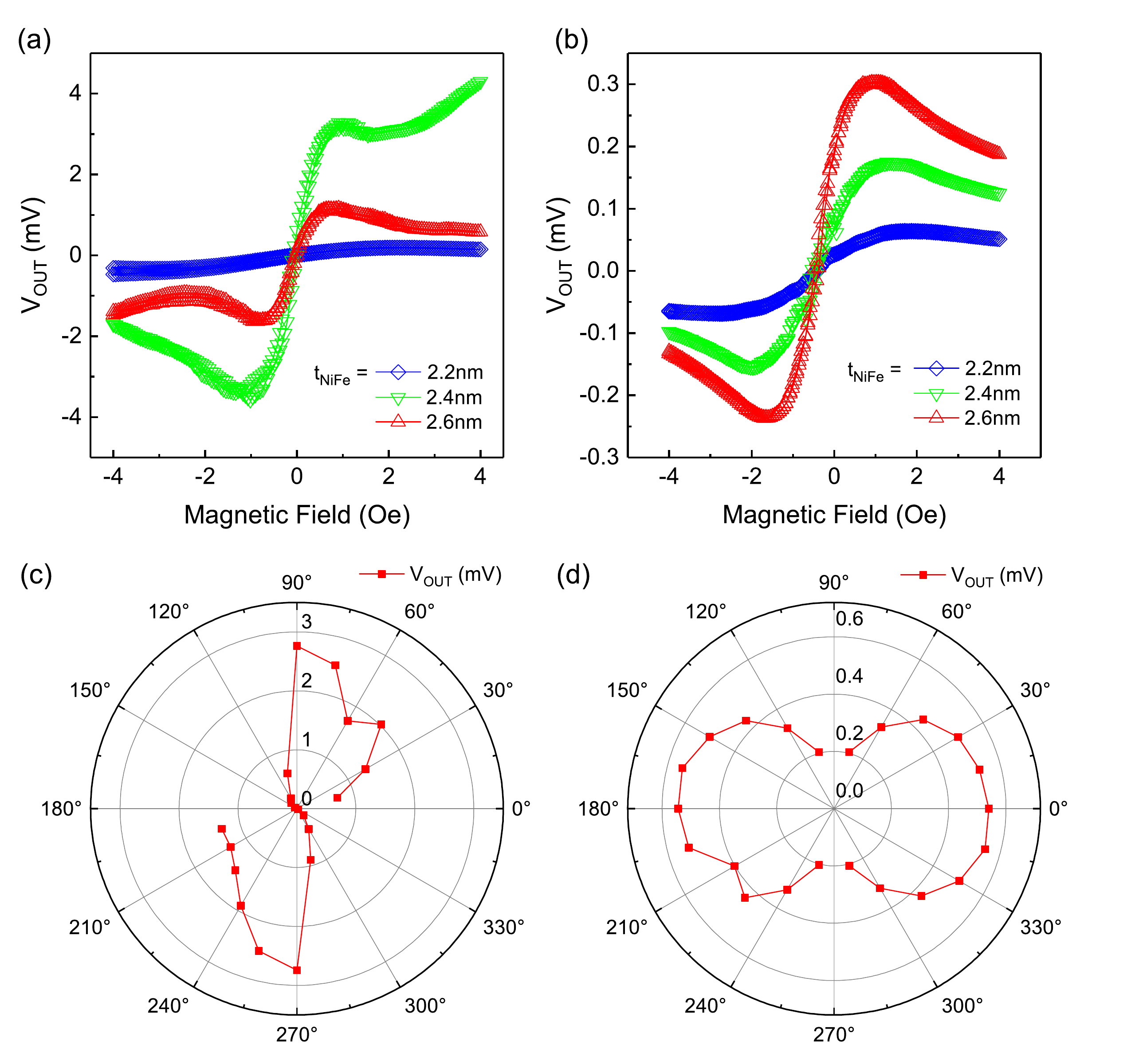}
\caption{\label{fig2} 
Thickness dependence of transfer characteristics of Pt (2\,nm)/NiFe (2.2, 2.4, 2.6\,nm) sensors under (a) DC excitation and (b) AC excitation; (c) and (d): angular plot of the amplitude of linear range for the Pt(2\,nm)/NiFe(2.6\,nm) sensor under DC (c) and AC (d)  excitations.
}
\end{figure}

Figure~\ref{fig2}(a) and \ref{fig2}(b) compare the transfer curves of sensors with different NiFe thicknesses: \( t_{\text{NiFe}} = 2.2 \), 2.4, and 2.6~nm. To facilitate comparison, we have subtracted the offset from the DC response curves, whereas the AC response curves are presented using the raw data. The AC response was obtained by applying the magnetic field along the \( x \)-axis, while the DC response was measured with the field applied along the \( y \)-axis. The DC current used was 3~mA for all samples. For the AC measurements, the same current amplitudes were used, with the excitation frequency fixed at 5 kHz. As shown in Figs.~\ref{fig2}(a) and \ref{fig2}(b), the AC response increases monotonically with increasing NiFe thickness, whereas the DC response reaches a peak at \( t_{\text{NiFe}} = 2.4 \, \text{nm} \). The low signal amplitude for the 2.2~nm sample is presumably due to its superparamagnetic behavior, which results in poor response to low magnetic field at room temperature. As the NiFe thickness increases, its magnetic properties improve, leading to enhanced AC response. In the case of the DC response, increasing the thickness to 2.6~nm may promote the formation of larger magnetic domains. Depending on the domain configuration, the sensitivity to externally applied fields can vary with direction, potentially explaining the non-monotonic behavior of the DC response with thickness. These results highlight the different mechanisms governing the AC and DC sensing modes and underscore the importance of optimizing magnetic layer thickness for specific sensing applications.

The amplitude of the DC signal is approximately one order of magnitude larger than that of the AC signal under the specific measurement condition, i.e., \( I = \, \text{3} \, \text{mA} \), with a field sensitivity of 38.22 m$\Omega$/Oe, 91.67 m$\Omega$/Oe under AC excitation and 1263 m$\Omega$/Oe, 622.22 m$\Omega$/Oe under DC excitation for \( t_{\text{NiFe}} = 2.4 \) nm and 2.6 nm respectively. Moreover, the DC response exhibits a significant offset, while the offset in the AC response is negligible. This observation is consistent with our previous findings on SMR sensors using elliptical sensing elements.\cite{xu2018ultrathin} The large DC offset arises from resistance imbalance between the sensor arms (e.g., AC and CD), which is likely introduced during the fabrication process. However, such offset is effectively suppressed in the second-harmonic (or equivalently AC) measurement, since it manifests itself as the first-harmonic component and is filtered out. The nearly zero offset is one of the key advantages of AC-driven sensors compared to their DC counterparts.

The angular dependence of the DC and AC responses of the Pt(2\,nm)/NiFe(2.4\,nm) sensor was investigated by varying the angle of the external magnetic field from \( 0^\circ \) to \( 360^\circ \) in steps of \( 15^\circ \), and the results (i.e., the linear amplitude) are shown in Figures~\ref{fig2}(c) and \ref{fig2}(d), respectively. As can be seen from Fig.~\ref{fig2}(c), under DC excitation, the linear response region appears in a narrow range near $\varphi_H$ = $90^\circ$. In contrast, under AC excitation, the sensor exhibits clear and symmetric linear regions across a much broader range of magnetic field angles. The response reaches its maximum at $\varphi_H = 0^\circ$ and $180^\circ$, gradually diminishes with increasing angular deviation, and eventually disappears around $\varphi_H = 90^\circ$ and $270^\circ$. These results clearly demonstrate the bi-axial sensing capability of the ring-shaped SMR sensor under different modes of excitations.

\begin{figure}[htbp]
\centering
\includegraphics[width=\linewidth]{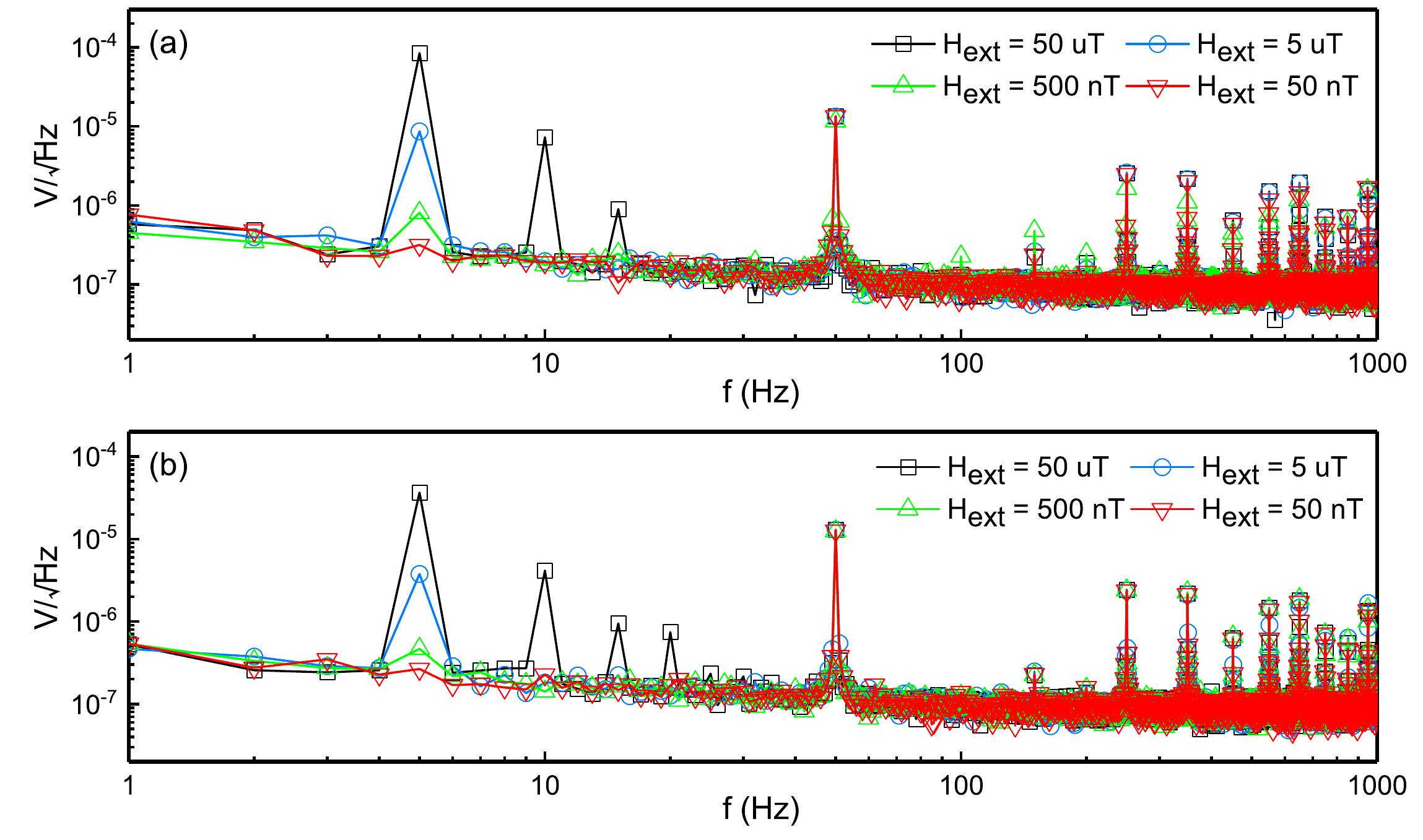}
\caption{\label{fig3}Noise spectral density of the $t_{\text{NiFe}} = 2.2$~nm sensor under DC (a) and AC (b) excitation. An AC excitation field with a fixed frequency of 5~Hz and varying amplitudes ($50$~nT - 50 $\mu${T}) was applied during the noise measurements. }
\end{figure}

\begin{figure}[htbp]
\centering
\includegraphics[width=\linewidth]{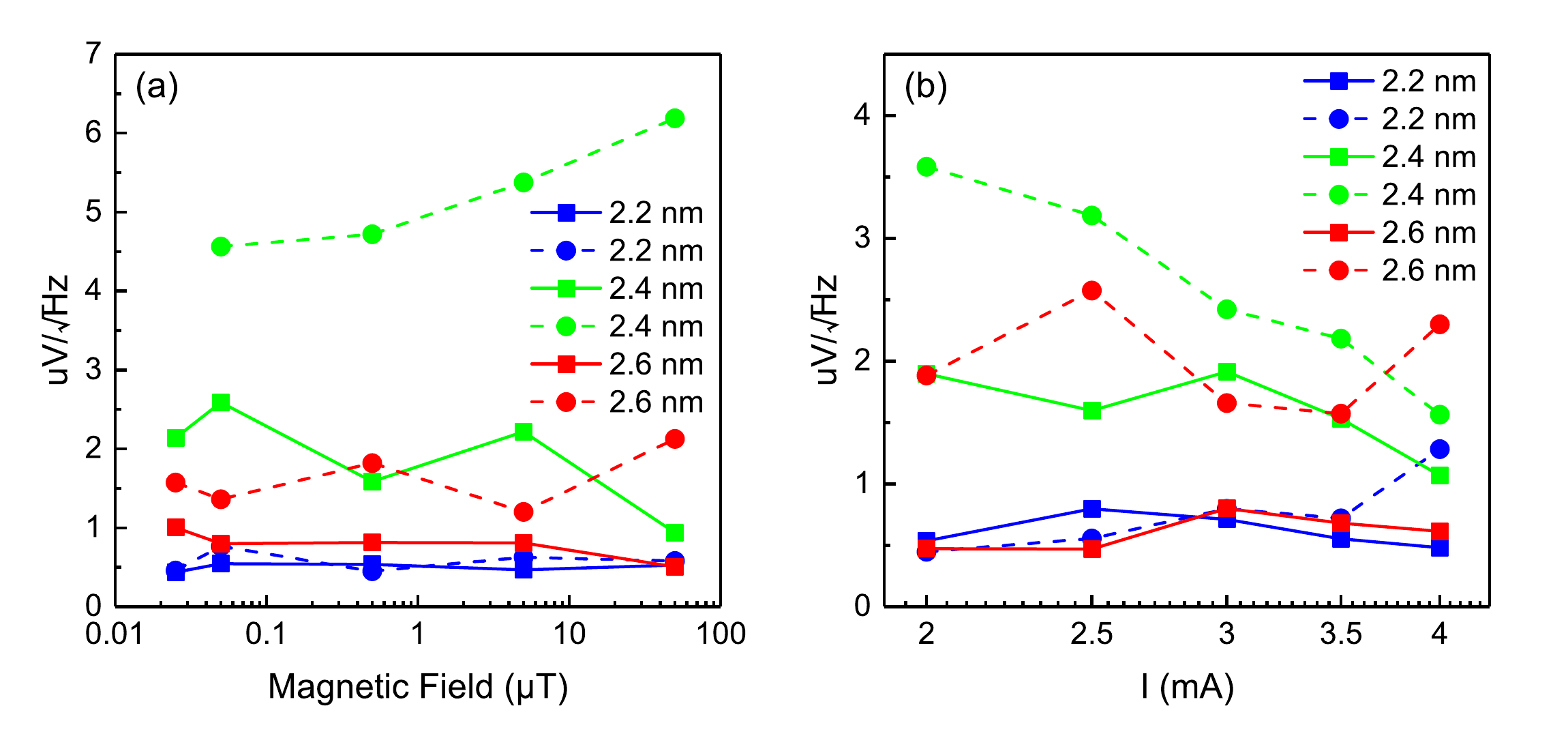}
\caption{\label{fig4}Noise spectral density of Pt (2\,nm)/NiFe sensors ($t_\mathrm{NiFe}$ = 2.2, 2.4, and 2.6\,nm) measured at 1\,Hz as a function of (a) peak AC field and (b) sensing current. Solid-lines correspond to AC excitation, whereas dotted-lines are for DC measurements. 
} 
\end{figure}

We now turn to the noise characteristics of the sensors operated in different modes, which were measured in a magnetically shielded environment using a dynamic signal analyzer. In order to assess the detection limit of the sensors, sinusoidal magnetic fields with a fixed frequency of 5~Hz but varying amplitudes were applied in the respective sensitive directions of the AC- and DC-operated sensors during the noise measurements. Figs.~\ref{fig3}(a) and \ref{fig3}(b) present the noise spectra of the sensor ($t_{\text{NiFe}} = 2.2$ nm) under DC and AC excitations, respectively. The sharp spikes at and above 50~Hz can be ignored as they are caused by the power line interference. The main peak at 5~Hz and the corresponding higher harmonics are due to the applied field, with the latter appearing only when the field exceeds the sensor’s dynamic range.  The amplitude of the peak decreases with reduced magnetic field strength, confirming that the sensor is capable of detecting external magnetic fields as low as 50 nT. At low frequencies, the noise is dominated by \( 1/f \) noise,\cite{lahav2024planar} which, in magnetic sensors, generally consists of a non-magnetic and a magnetic component. The non-magnetic component originates from resistance fluctuations, while the magnetic component is commonly attributed to thermally induced domain wall nucleation and motion. This magnetic noise can be phenomenologically described as follows:\cite{grosz2013planar,guo2015reduction}

\begin{equation}
S_V = \frac{\delta_H \, V_b^2}{N_c \cdot \mathrm{Vol} \cdot f}
\end{equation}

\noindent where \( \delta_H \) is the Hooge constant, \( V_b \) is the bias voltage across the bridge, \( N_c \) is the free electron density, \( Vol \) is the effective volume of the NiFe layer, and \( f \) is the frequency. The Hooge constant \( \delta_H \) characterizes the amplitude of the \( 1/f \) noise. As shown in our previous work,\cite{xu2018ultrathin} the magnetic contribution to \( \delta_H \) can be significantly reduced by continuously flipping the magnetization direction. Although the effective field in sensor operation may not be strong enough to induce full switching, the oscillation of magnetization around its equilibrium position due to the AC effective field is expected to contribute to the observed noise reduction. If we let \( V_b = 1 \, \text{V} \), \( f = 1 \, \text{Hz} \), \( N_c = 1.5 \times 10^{29} \, \text{m}^{-3} \) for NiFe, \( \delta_H = 0.001 \) to \( 0.1 \), and \( Vol \) be the volume of a ring with outer radius of 150 \(\mu\text{m}\), width 10 \(\mu\text{m}\), and thickness 2.4 nm, \( \sqrt{S_V} \) turned out to be \(\sim 1.74 \times 10^{-8} - 1.74 \times 10^{-6} \, \text{V}/\sqrt{\text{Hz}}\), which is roughly in the same range of the measurement results.\\
\indent To quantify the noise performance in different operation modes, we show in Fig.~\ref{fig4} the noise amplitude density at 1~Hz for AC-operated (solid lines) and DC-operated (dotted lines) sensors as a function of (a) the amplitude of the applied AC field and (b) the sensing current. Except for the sensor with \( t_{\text{NiFe}} = 2.2~\text{nm} \), the noise is consistently much lower in AC-operated sensors compared to their DC counterparts. Moreover, the noise generally decreases with increasing sensing current, consistent with previous findings on SMR sensors employing elliptical elements.\cite{xu2018ultrathin} However, the noise levels observed in this study are notably higher than those reported for elliptical SMR sensors. This discrepancy is presumably due to the fact that, although a magnetic field along the \( x \)-axis was applied during deposition to induce uniaxial anisotropy, the actual easy axis may have a non-uniform distribution around the ring-shaped structure due to shape-induced anisotropy. This variation in magnetic anisotropy can result in instability of magnetic domains during sensor operation, thereby increasing noise. One potential approach to mitigating this issue is to introduce a soft exchange bias, which stabilizes the domain configuration throughout the ring while simultaneously providing the SOT effective field. This strategy, along with other methods for improving noise performance, will be explored in future studies.\\
\indent In summary, we have devised a ring-shaped SMR sensor capable of detecting magnetic fields along two perpendicular directions by simply operating it under either AC or DC excitation. Although the AC-mode produces a weaker output signal compared to the DC-mode, it exhibits significantly lower noise levels, resulting in comparable or even better performance in low-field magnetic detection. In its current form, the sensor is capable of detecting magnetic fields as low as approximately 50~nT. We anticipate that the detectivity can be significantly improved once the magnetic domain structures are stabilized, for instance by introducing exchange bias. Such sensors with biaxial sensing capability hold promise for a wide range of applications, including vector magnetic field detection, wearable biomedical devices, navigation systems, and compact geomagnetic sensing platforms.

Y.W. acknowledges the funding supported by RIE2025 Manufacturing, Trade and Connectivity (MTC) Programmatic Fund, Agency for Science, Technology and Research under Grant No. M24N9b0117. J.X. acknowledge the support of scholarship from NUS Guangzhou Research Translation and Innovation Institute (NUS GRTII).

\section*{Author Declarations}

\subsection*{Conflict of Interest}
The authors have no conflicts to disclose.

\subsection*{DATA AVAILABILITY}
The data that support the findings of this study are available from the corresponding author upon reasonable request.

\subsection*{REFERENCES}
\bibliography{reference}

\end{document}